\documentclass[twocolumn,showpacs,reprint,preprintnumbers,amsmath,amssymb,apl]{revtex4-1}
\usepackage{amsmath}
\usepackage{txfonts}
\usepackage{graphicx}
\usepackage{dcolumn}
\usepackage{bm}
\usepackage{color}
\usepackage[colorlinks=true,citecolor=blue,linkcolor=magenta]{hyperref}

\begin{document}

\title{Radiation pressure induced difference-sideband generation beyond linearized description}
\author{Hao Xiong}\email{haoxiong1217@gmail.com}
\author{Yu-Wan Fan}
\author{Xiaoxue Yang}
\author{Ying Wu}\email{yingwu2@126.com}
\affiliation{School of Physics, Huazhong University of Science and Technology, Wuhan, 430074, P. R. China}
\date{\today}

\begin{abstract}
We investigate radiation-pressure induced generation of the frequency components at the difference-sideband in an optomechanical system, which beyond the conventional linearized description of optomechanical interactions between cavity fields and the mechanical oscillation. We analytically calculate amplitudes of these signals, and identify a simple square-root law for both the upper and lower difference-sideband generation which can describe the dependence of the intensities of these signals on the pump power. Further calculation shows that difference-sideband generation can be greatly enhanced via achieving the matching conditions. The effect of difference-sideband generation, which may have potential application for manipulation of light, is especially suited for on-chip optomechanical devices, where nonlinear optomechanical interaction in the weak coupling regime is within current experimental reach.
\end{abstract}

\maketitle

Resonantly enhanced feedback-backaction of optomechanical coupling \cite{rev} has attracted great interest recently and the strong interaction between cavity fields and mechanical motion has been demonstrated experimentally \cite{exp}. This emerging subject leads to many potential applications for both optics and physics, including achieving high precision measurement \cite{measurement}, on-chip manipulation of asymmetric light propagation \cite{nonreciprocity,nonreciprocity2}, and optomechanically induced transparency \cite{omit,omit1,omit2}. Optomechanically induced transparency is an interesting analog of electromagnetically induced transparency, where the control field induces a transmission window for the probe field when the resonance condition is met \cite{omit3,omit4,omit5}. Optomechanically induced transparency can be well understood through the linearization of the semiclassical evolution equations \cite{omit6,omke1}.


In view of the nonlinear nature of the interaction between light and mechanical motion via radiation pressure, many interesting phenomena and applications have been revealed. A perturbative analysis of output optical spectrum in the parameter configuration of optomechanically induced transparency reveals spectral components at the second order sideband that arises from nonlinear optomechanical interactions and exhibits a prominent feature of nonlinear optomechanically induced transparency \cite{1}. Recently, nonlinear optomechanical dynamics have emerged as an interesting frontier in cavity optomechanics \cite{om,hsg,hx}. In the semiclassical mechanism, sideband generation and optomechanical chaos have been studied in various contents, including coherent-mechanical pumped optomechanical systems \cite{mechanicalpump}, optomechanical system with second-order coupling \cite{nonlinearcoupling}, hybrid electro-optomechanical systems \cite{heom,heom2}, and photonic molecule optomechanical system \cite{photonicmolecule}. Delaying or advancing higher-order sideband signals \cite{delaying} have also been revealed which may be important in optical information processing techniques.

\begin{figure}[ht]
\centering
\includegraphics[width=0.45\textwidth]{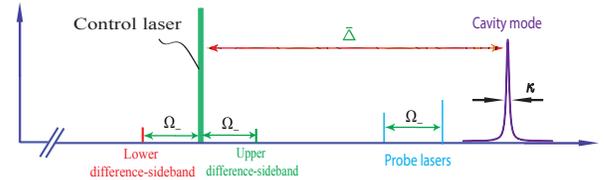}
\caption{\label{fig:0} Frequency spectrogram of difference-sideband generation in an optomechanical system with two probe fields.}
\end{figure}

In the present work, we consider that the optomechanical system is double driven: two probe fields with different frequencies ($\omega_1$ and $\omega_2$, respectively) perturb the steady state cavity field provided by a strong control field with the frequency $\omega_c$. Generation of spectral components at difference sideband [with frequency $\pm(\omega_1-\omega_2)$ in a frame rotating at $\omega_c$, as shown in Fig. \ref{fig:0}] is demonstrated analytically.  We find a simple square-root law which can describe the dependence of difference-sideband generation on the pump power. Our results also reveal some interesting matching conditions that difference-sideband generation becomes efficient when one of the matching conditions is met. To explain the physical interpretation of these matching conditions, features of the mechanical oscillation at difference sideband are also discussed.

The signals at the difference sideband may be important in understanding the nonlinear optomechanical interactions, where nonlinear features of optomechanical systems with multiple probe field driven is still unknown. From the precision measurement perspective, matching conditions of the difference sideband may provide an potential method for determination of parameters \cite{measurement} and phonon number \cite{nonlinear} of optomechanical systems. In addition, a robust difference-sideband generation that works under low operating power may be useful for optical information processing, and the effect of difference-sideband generation provides an effective way to manipulate light in a solid-state architecture. The present mechanism of difference-sideband generation may also be applied to other similar systems, such as quantum dot and well system \cite{well,well2}, metasurfaces \cite{metasurfaces}, graphene \cite{graphene}, and even DNA-quantum dot hybrid system \cite{DNA}.

We consider that the optomechanical system, which formed by a fixed mirror and a movable mirror with effective mass $m$ and angular frequency $\Omega_m$, is driven by a strong control field with the frequency $\omega_c$ and two probe fields with frequencies $\omega_1$ and $\omega_2$. In the parameter configuration of optomechanically induced transparency, the frequency of the control field is detuned by $\bar{\Delta}\approx -\Omega_m$ from the cavity resonance frequency, and the Hamiltonian formulation of the optomechanical system is \cite{omit}:
\begin{gather}
\hat{H}=\frac{\hat{p}^2}{2m}+\frac{m\Omega_m^2\hat{x}^2}{2}+\hbar \omega_0 \hat{a}^\dag \hat{a} -\lambda_0 \hat{x} \hat{a}^\dag \hat{a} +\hat{H}_{\mathrm{control}}+\hat{H}_{\mathrm{probe}},\nonumber
\end{gather}
where $\hat{H}_{\mathrm{control}}=\mathrm{i}\hbar\sqrt{\eta\kappa} \varepsilon_c (\hat{a}^\dag e^{-\mathrm{i}\omega_c t}- \hat{a} e^{\mathrm{i}\omega_c t})$ and $\hat{H}_{\mathrm{probe}}=\mathrm{i}\hbar \sqrt{\eta\kappa}(\hat{a}^\dag\varepsilon_1 e^{-\mathrm{i}\omega_1 t}+ \hat{a}^\dag\varepsilon_2 e^{-\mathrm{i}\omega_2 t}- \mathrm{H.c.})$, $\hat{p}$ ($\hat{x}$) is the momentum (position) operator of the movable mirror. $\omega_0$ is the resonance frequency of the cavity, $\hat{a}$ ($\hat{a}^\dag$) is the annihilation (creation) operator of the cavity field with line width $\kappa$ in the resolved-sideband regime, $\lambda_0=-\hbar G$ with $G$ the optomechanical coupling constant \cite{Law}. The coupling parameter $\eta$ is chosen to be the critical coupling value $1/2$ \cite{omit} here, and $\varepsilon_i=\sqrt{P_i/\hbar \omega_i}$ ($i$=c, 1, 2) are the amplitudes of the input fields with $P_c$ the pump power of the control field and $P_1$ ($P_2$) the power of the probe field with frequency $\omega_1$ ($\omega_2$).

Based on the Hamiltonian, the intracavity field and the mechanical displacement can be described by the Heisenberg equations (in a frame rotating at $\omega_c$):
\begin{gather}
\dot{a}=[\mathrm{i}(\Delta+\lambda_0 x/\hbar)-\kappa]a+\sqrt{\eta\kappa}(\varepsilon_c+s_{\mathrm{in}}),\label{equ:2.1}\\
\biggl(m\frac{\mathrm{d}^2}{\mathrm{d}t^2}+ m\Gamma_m \frac{\mathrm{d}}{\mathrm{d}t} + m\Omega_m^2\biggr) x=\lambda_0 a^* a,\label{equ:2.2}
\end{gather}
where the operators are reduced to their expectation values, viz. $a(t)\equiv \langle \hat{a}(t)\rangle$ and $x(t)\equiv \langle \hat{x}(t)\rangle$, the mean-field approximation by factorizing averages is used and the quantum noise terms are dropped. $\Delta=\omega_c-\omega_0$, $s_{\mathrm{in}}=\varepsilon_1 e^{-\mathrm{i}\delta_1 t}+\varepsilon_2  e^{-\mathrm{i}\delta_2 t}$ with $\delta_1=\omega_1-\omega_c$ and $\delta_2=\omega_2-\omega_c$, and $\Gamma_m$ is the decay rate of the mechanical oscillator. The solution of Eqs. (\ref{equ:2.1}) and (\ref{equ:2.2}) can be written as $a=\bar{a}+\delta a$ and $x=\bar{x}+\delta x$, where $\bar{a}=\sqrt{\eta\kappa}\varepsilon_c/(-\mathrm{i}\bar{\Delta}+\kappa)$ and $\bar{x}=\lambda_0 |\bar{a}|^2/(m\Omega_m^2)$, with $\bar{\Delta}=\Delta+\lambda_0 \bar{x}/\hbar$, and $\delta a$ and $\delta x$ obey the following equations:
\begin{gather}
\frac{\mathrm{d}}{\mathrm{d}t}\delta a=(\mathrm{i}\bar{\Delta}- \kappa)\delta a +\mathrm{i} \lambda_0 (\bar{a} \delta x + \delta x\delta a)/\hbar+\sqrt{\eta\kappa}s_{\mathrm{in}},\nonumber\\
\hat{\Psi} \delta x= \lambda_0 (\bar{a}\delta a^*+\bar{a}^*\delta a+\delta a^*\delta a),\label{equ:delta_equation}
\end{gather}
where $\hat{\Psi}=m(\mathrm{d}^2/\mathrm{d}t^2+ \Gamma_m \mathrm{d}/\mathrm{d}t + \Omega_m^2)$. By neglecting the nonlinear terms $\mathrm{i} \lambda_0 \delta x\delta a$ and $\lambda_0 \delta a^*\delta a$, Eqs. (\ref{equ:delta_equation}) can be solved analytically with the ansatz $\delta a^{L} =a_{\delta_1}^+e^{-\mathrm{i}\delta_1t}+a_{\delta_1}^-e^{\mathrm{i}\delta_1t} +a_{\delta_2}^+e^{-\mathrm{i}\delta_2t} +a_{\delta_2}^-e^{\mathrm{i}\delta_2t}$ and $\delta x^{L}=x_{\delta_1} e^{-\mathrm{i}\delta_1t}+x_{\delta_1}^* e^{\mathrm{i}\delta_1t}+x_{\delta_2} e^{-\mathrm{i}\delta_2t}+x_{\delta_2}^* e^{\mathrm{i}\delta_2t}$, where the frequency space is closed. While such linearized dynamics can explain many phenomena arise in cavity optomechanics, the nonlinear terms $\mathrm{i} \lambda_0 \delta x\delta a$ and $\lambda_0 \delta a^*\delta a$ must be taken account for discussion of difference-sideband generation, which is out of the frequency space of linearized dynamics.

Before passing to the process of constructing solutions, we examine the nonlinear term in Eqs. (\ref{equ:delta_equation}). The nonlinear term (here we take $\delta x\delta a$ as an example, the another nonlinear term $\delta a^*\delta a$ is similar) approximated by linear solutions consists of contributions at various frequencies:
\begin{gather}
\delta x^{L} \delta a^{L}={B_0} + {x_{{\delta _1}}}a_{{\delta _1}}^ + {e^{ - 2\mathrm{i}{\delta _1}t}} + x_{{\delta _1}}^*a_{{\delta _1}}^ - {e^{{2\mathrm{i}}{\delta _1}t}} \qquad\qquad\qquad\nonumber\\
\qquad + {x_{{\delta _2}}}a_{{\delta _2}}^ + {e^{ - 2{\text{i}}{\delta _2}t}}
+ x_{{\delta _2}}^*a_{{\delta _2}}^ - {e^{{2\mathrm{i}}{\delta _2}t}}
+ B_1{e^{ - {\text{i(}}{\delta _1} + {\delta _2})t}} \nonumber\\
\qquad+ B_2{e^{ - {\text{i(}}{\delta _1} - {\delta _2})t}}
+ B_3{e^{{\text{i(}}{\delta _1} - {\delta _2})t}}
+ B_4{e^{{\text{i(}}{\delta _1} + {\delta _2})t}},
\end{gather}
where $B_0={x_{{\delta _1}}}a_{{\delta _1}}^ -  + x_{{\delta _1}}^*a_{{\delta _1}}^ +  + {x_{{\delta _2}}}a_{{\delta _2}}^ -  + x_{{\delta _2}}^*a_{{\delta _2}}^ +$, $B_1={x_{{\delta _1}}}a_{{\delta _2}}^ + + {x_{{\delta _2}}}a_{{\delta _1}}^ + $, $B_2={x_{{\delta _1}}}a_{{\delta _2}}^ -  + x_{{\delta _2}}^*a_{{\delta _1}}^ + $, $B_3=x_{{\delta _1}}^*a_{{\delta _2}}^ +  + {x_{{\delta _2}}}a_{{\delta _1}}^ - $, $B_4=x_{{\delta _1}}^*a_{{\delta _2}}^ -  + x_{{\delta _2}}^*a_{{\delta _1}}^ -$. The frequency components of $\pm(\delta_1+\delta_2)$ and $\pm(\delta_1-\delta_2)$ are called sum- and difference- sideband. The physical picture of such process due to the nonlinear terms $\mathrm{i} \lambda_0 \delta x\delta a/\hbar$ and $\lambda_0\delta a^*\delta a$ is very similar to sum- and difference- frequency generation in a nonlinear medium \cite{sd}. Both sum- and difference-sidebands have analogous dependence on the pump power and the frequencies of the two probes. Here we focus only on the difference-sidebands due to lacking of space. A full treatment of difference-sideband generation in the perturbative regime can be performed by introducing the nonlinear ansatz: $\delta a = a_1^+e^{-\mathrm{i}\delta_1t} +a_1^-e^{\mathrm{i}\delta_1t} +a_2^+e^{-\mathrm{i}\delta_2t} +a_2^-e^{\mathrm{i}\delta_2t} +a_d^+e^{-\mathrm{i}\Omega_- t}+a_d^-e^{\mathrm{i}\Omega_- t}+\cdots$ and $\delta x=x_1 e^{-\mathrm{i}\delta_1t} +x_1^* e^{\mathrm{i}\delta_1t} +x_2 e^{-\mathrm{i}\delta_2t} +x_2^* e^{\mathrm{i}\delta_2t} +x_d e^{-\mathrm{i}\Omega_- t} +x_d^* e^{\mathrm{i}\Omega_- t}+\cdots$, with $\Omega_-=\delta_1-\delta_2$. Other frequency components, including second- and higher-order sidebands \cite{1}, are ignored due to the fact that these components contribute little to difference-sideband generation in the perturbative regime.

Substitution of the nonlinear ansatz into Eqs. (\ref{equ:delta_equation}) leads to there matrix equations \cite{1}: $M({\delta _1}){\alpha _1} = {\beta _1}$, $M({\delta _2}){\alpha _2} = {\beta _2}$, and $M({\Omega_-}){\alpha _d} = {\beta _d}$, where $\alpha_i=[{a_i^ + }, {{{(a_i^ - )}^*}}, {{x_i}}]^T $ with $i= 1, 2, d$, $\beta_1=[{\sqrt{\eta\kappa}{\varepsilon_1}}, 0, 0]^T$, $\beta_2=[{\sqrt {\eta\kappa}{\varepsilon_2}}, 0, 0]^T$,
\begin{gather}
\beta_d=\frac{{{\text{i}}{\lambda _0}}}
{\hbar }\left( {\begin{array}{*{20}{c}}
   {a_1^ + x_2^* + a_2^ - {x_1}}  \\
   { - {{(a_1^ - )}^*}x_2^* - {{(a_2^ + )}^*}{x_1}}  \\
   { - {\text{i}}\hbar [a_1^ + {{(a_2^ + )}^*} + a_2^ - {{(a_1^ - )}^*}]}  \\
 \end{array} } \right),\nonumber\\
M(x) =
\left(
  \begin{array}{ccc}
   \theta (-x) & 0 & -\text{i}\lambda_0 \bar a /\hbar \\
   0 & [\theta{(x)}]^* & \text{i}\lambda_0 {\bar a}^* /\hbar \\
   { - {\lambda _0}{{\bar a}^*}} & { - {\lambda _0}\bar a} & {\sigma (x)}  \\
  \end{array}
\right),
\end{gather}
with $\theta (x) = s + \mathrm{i}x$, $\sigma (x) = m\Omega _m^2 - mx^2 - \mathrm{i}m{\Gamma _m}x$, and $s=\kappa -\mathrm{i}\bar \Delta$. The solution to these equations can be obtained as follows:
\begin{gather}
a_1^ +  =\frac{{{\sqrt{\eta\kappa}\varepsilon _1}\tau ({\delta _1})}}{{\theta ( - {\delta _1})\tau ({\delta _1}) - \alpha }},\quad
{x_1} = \frac{{{\lambda _0}{{\bar a}^*}a_1^ + }}{{\tau ({\delta _1})}},\quad
a_1^ -  = \frac{{\mathrm{i}{\lambda _0}\bar a}}{{\hbar \theta ({\delta _1})}}x_1^*,\nonumber\\
a_2^ +  =\frac{{{\sqrt{\eta\kappa}\varepsilon _2}\tau ({\delta _2})}}{{\theta ( - {\delta _2})\tau ({\delta _2}) - \alpha }},\quad
{x_2} = \frac{{{\lambda _0}{{\bar a}^*}a_2^ + }}{{\tau ({\delta _2})}},\quad
a_2^ -  = \mathrm{i}\frac{{{\lambda _0}\bar a}}{{\hbar \theta ({\delta _2})}}x_2^*,\nonumber\\
a_d^ +  = \mathrm{i}\frac{{{\lambda _0}}}{\hbar }\frac{{{\lambda _0}\bar a{\xi _d} + (a_1^ + x_2^* + a_2^ - {x_1})\tau ({\Omega _ - })}}
{{\tau ({\Omega _ - })\theta ( - {\Omega _ - }) - \alpha }},\nonumber\\
a_d^ - =\frac{{\mathrm{i}{\lambda _0}(\bar ax_d^*{\text{ + }}a_1^ - {x_2} + a_2^ + x_1^*)}}{{\hbar \theta ({\Omega _ - })}}, \quad
x_d = \frac{{{\lambda _0}({\xi _d} + {{\bar a}^*}a_d^ + )}}{{\tau ({\Omega _ - })}},\label{equ:solution}
\end{gather}
where $\alpha=\mathrm{i}{\lambda _0}^2{|\bar a|}^2/\hbar$, $\tau (x) = \sigma (x) + \alpha/{\theta {{(x)}^*}}$, ${\xi _d} = a_1^ + {(a_2^ + )^*} + a_2^ - {(a_1^ - )^*} - \mathrm{i}{\lambda _0}\bar a[{(a_1^ - )^*}x_2^* + {(a_2^ + )^*}{x_1}]/[\hbar \theta {(\Omega_-)}^*]$.

Using the input-output relation $s_{\mathrm{out}}=s_{\mathrm{in}}-\sqrt{\eta \kappa }a$, the amplitude of the output field at upper and lower difference sideband can be obtained as $-\sqrt{\eta\kappa}a_d^+$ and $-\sqrt{\eta\kappa}a_d^-$, respectively. We define $\eta_d^+=|-\sqrt{\eta \kappa}a_d^+/\varepsilon_1|$, which is the ratio between the amplitude of the output field at the upper difference sideband and the amplitude of the first input probe field, as the efficiency of the upper difference-sideband generation process. Here the denominator (the amplitude of the first input probe field) $\varepsilon_1$ is just chosen for convenience, and therefore leads to the efficiency being dimensionless. One also can define the amplitude of the second input probe field $\varepsilon_2$ as the denominator. Similarly, $\eta_d^-=|-\sqrt{\eta\kappa}a_d^- /\varepsilon_1|$ is the efficiency of the lower difference-sideband generation process.

There is a high dependence of the efficiencies of difference-sideband generation on the pump power of the control field $P_c$ and the frequencies of the probe fields. For the relationship between the efficiencies of difference-sideband generation and the pump power of the control field $P_c$, we find a simple square-root law which can describe $\eta_d^+$ ($\eta_d^-$) vary with $P_c$ well. For the dependence of difference-sideband generation on the frequencies of the probe fields, we identify the matching conditions of both upper and lower difference-sideband generation, where the efficiencies are enhanced significantly when the matching conditions are met.

\begin{figure}[ht]
\centering
\includegraphics[width=0.36\textwidth]{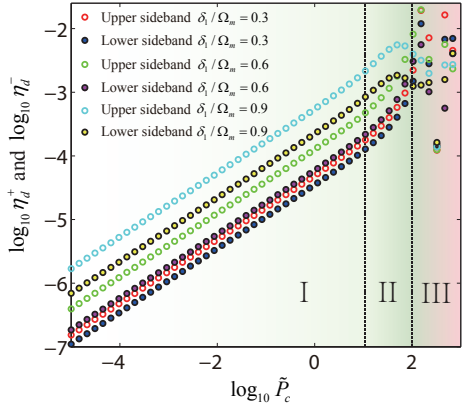}
\caption{\label{fig:1}(Color online) Dependencies of the efficiencies of difference-sideband generation on the pump power of the control field. The parameters used in the calculation are \cite{omit} $m$=20 ng, $G/2\pi$=-12 GHz/nm, $\Gamma_m/2\pi$=41.0 kHz, $\kappa/2\pi$=15.0 MHz, $\Omega_m/2\pi$=51.8MHz, and $\Delta$=$-\Omega_m$. The wavelength of the control field is chosen to be 532 nm here, $\delta_2=0.1 \Omega_m$, $P_0$=1 mW, and $P_1=P_2=10$ $\mu$W.}
\end{figure}

In the resolved-sideband regime, the lower sideband which is far off-resonance can be neglected. In addition, by considering the fact that $\lambda_0\ll \hbar\Omega_m/\bar{x}$, we can simplify the solutions of $a_d^+$ and $a_d^-$ as follows:
\begin{gather}
a_d^ +  = \Theta_1 \sqrt {\tilde{P}_c}, \qquad  a_d^ -  = \Theta_2 \sqrt {\tilde{P}_c},  \nonumber\\
\Theta_1  = \frac{{{\text{i}}\lambda _0^2a_1^ + {{(a_2^ + )}^*}}}
{{\hbar s\theta ( - {\Omega _ - })}}\left( {\frac{1}
{{\tau ({\Omega _ - })}} + \frac{1}
{{{\tau ^*}({\delta _2})}}} \right)\sqrt {\frac{{\eta \kappa }}
{{\hbar {\omega _c}}}{P_0}},\nonumber\\
\Theta_2=\frac{{{\text{i}}\lambda _0^2a_2^ + {{(a_1^ + )}^*}}}
{{\hbar s\theta ({\Omega _ - })}}\left( {\frac{1}
{{{\tau ^*}({\Omega _ - })}} + \frac{1}
{{{\tau ^*}({\delta _1})}}} \right)\sqrt {\frac{{\eta \kappa }}
{{\hbar {\omega _c}}}{P_0}},\label{equ:sqrtlaw}
\end{gather}
where $\tilde{P}_c=P_c/P_0$ with $P_0$ an arbitrary power to make $\tilde{P}_c$ dimensionless, and $\Theta_1$ and $\Theta_2$ are almost independent of the power $\tilde{P}_c$. Such square-root law for difference-sideband generation is quite accurate when $P_c$ is not high enough. Figure \ref{fig:1} shows (in logarithmic form) the dependencies of the efficiencies of difference-sideband generation on the pump power of the control field. The linear relation between ${\log _{10}}\eta _d^ + $ (${\log _{10}}\eta _d^ - $) and ${\log _{10}}\left( {{{\tilde P}_c}} \right)$ confirms the square-root law (\ref{equ:sqrtlaw}), which holds, for both upper and lower difference-sideband generation, until $\log_{10}\tilde{P}_c=1$ (equally $P_c=10$ mW) in the region I. When the pump power of the control field reaches region II, deviation from the linear relation between ${\log _{10}}\eta _d^ + $ (${\log _{10}}\eta _d^ - $) and ${\log _{10}}\left( {{{\tilde P}_c}} \right)$ implies that the square-root law needs to be modified in this case. This system has bistability if the pump power of the control field is strong enough \cite{1,bistability}, and the square-root law (which corresponds to perturbative description) breaks down completely when the bistability occur (shown in region III). In fact, the data in Region III, which is calculated with Eq. (\ref{equ:solution}) perturbatively, is not valid and only indicate the destruction of the perturbative description.

\begin{figure}[ht]
\centering
\includegraphics[width=0.40\textwidth]{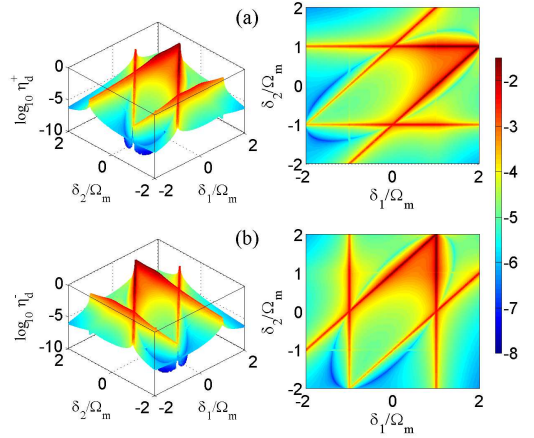}
\caption{\label{fig:2}(Color online) Efficiencies (in logarithmic form) of (a) upper difference-sideband generation and (b) lower difference-sideband generation versus $\delta_1$ and $\delta_2$. The parameters are the same as Fig. \ref{fig:1}.}
\end{figure}

To describe the dependence of difference-sideband generation on the the frequencies of the probe fields, calculation results of efficiencies (in logarithmic form) of difference-sideband generation as functions of both $\delta_1$ and $\delta_2$ are shown in Fig. \ref{fig:2}, where the efficiencies of difference-sideband generation exhibit peak structure for some specific values of $\delta_1$ and $\delta_2$. The processes of difference-sideband generation can be enhanced significantly through the suitable selection of $\delta_1$ and $\delta_2$. We call the specific values of $\delta_1$ ($\delta_2$) corresponding to these peaks as the matching conditions. From Fig. \ref{fig:2} one can identify the matching conditions for difference-sideband generation. Figure \ref{fig:2}(a) shows the calculation results of upper difference-sideband generation. One of matching conditions can be identified as $\delta_2\rightarrow \pm\Omega_m$, where the efficiency of upper difference-sideband generation is enhanced more significantly when $\delta_2\rightarrow \Omega_m$ than the case of $\delta_2\rightarrow -\Omega_m$, especially in the region around $\delta_1=\Omega_m$. In Fig. \ref{fig:2}(b), where the calculation results of efficiency of lower difference-sideband generation is shown, we observe a matching condition as $\delta_1\rightarrow \pm\Omega_m$. Both upper and lower difference-sideband generation are enhanced when $\delta_1-\delta_2\rightarrow \Omega_m$ which is the common matching condition for difference-sideband generation. Careful examination confirms that both Fig. \ref{fig:2}(a) and Fig. \ref{fig:2}(b) are not symmetrical for $\delta_1$ and $\delta_2$ due to the unequal status of the parameters $\delta_1$ and $\delta_2$ in the difference-sideband generation.

\begin{figure}[ht]
\centering
\includegraphics[width=0.36\textwidth]{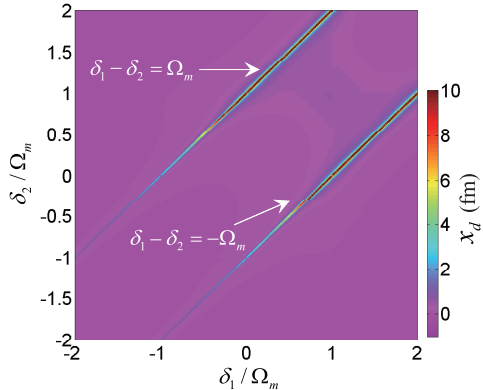}
\caption{\label{fig:3}(Color online) The amplitude of the mechanical oscillation at the difference-sideband in unit of femtometer vary with $\delta_1$ and $\delta_2$. The parameters are the same as Fig. \ref{fig:1}.}
\end{figure}

The physical interpretation of the common matching condition for difference-sideband generation relies on the the features of the mechanical oscillation at the difference sideband. Dependencies of the amplitude of the mechanical oscillation at the difference sideband on the variables $\delta_1$ and $\delta_2$ are illustrated in Fig. \ref{fig:3}. It is shown that $x_d$ becomes remarkable on the lines $\delta_1-\delta_2=\pm\Omega_m$, which are exactly corresponding to the common matching conditions for difference-sideband generation. The common matching condition can be understood as due to the beating between the two probe fields generates radiation pressure at the mechanical resonance frequency, hence excite significant mechanical oscillation. The special matching conditions $\delta_1= \pm\Omega_m$ and $\delta_2= \pm\Omega_m$ can also be understood through the resonance features of the mechanical oscillation, where the mechanical oscillation becomes significant if the beating between the control field and one of probe fields generates radiation pressure at the mechanical resonance frequency, and consequently leads to remarkable signals at the difference sideband via Stokes optomechanical scattering of the cavity fields.


In summary, by analyzing nonlinear optomechanical processes driven by double probe fields, we have demonstrated analytically difference-sideband generation in an optomechanical system, with the signal amplitude determined by the matching conditions and can be observed in the experimentally available parameter range. Further calculation shows that difference-sideband generation can be well controlled via adjusting the pump power of the control field. The effect of difference-sideband generation, which may offer insight into the understanding of optomechanical system and find applications in manipulation of light, is especially suited for on-chip optomechanical devices.

The work was supported by the National Basic Research Program of China (Grant No. 2016YFA0301200), the National Fundamental Research Program of China (Grant No. 2012CB922103), the National Science Foundation (NSF) of  China (Grant Nos. 11375067, 11275074, 11405061, and 11574104), and the Fundamental Research Funds for the Central Universities HUST (Grant No. 2014QN193).

\end{document}